\documentclass[aps,preprint,prd,floats,tightenlines,showpacs,showkeys,superscriptaddress,nofootinbib]{revtex4-1}

\usepackage[dvips]{graphicx,color}
\usepackage{amsmath,amssymb}
\usepackage{bm}
\usepackage{dcolumn}
\usepackage{epsfig}
\usepackage{natbib}

\usepackage{ulem}
\usepackage{color}

\definecolor{ggreen}{rgb}{0.2,0.6,0.2}

\newcommand{\ket}[1]{|{#1}\rangle}
\newcommand{\bra}[1]{\langle{#1}|}
\newcommand{\inp}[2]{\langle{#1}|{#2}\rangle}

\newcommand{\Slash}[1]{\ooalign{\hfil/\hfil\crcr$#1$}}
\def\braketa#1{\langle#1\rangle}

\begin{document}
\bibliographystyle{apsrev4-1}

\title{Extraction of  Meson Resonances
from Three-pions Photo-production Reactions
}
\author{S. X. Nakamura}
\affiliation{Yukawa Institute for Theoretical Physics, Kyoto University,
Kyoto 606-8502, Japan}
\affiliation{
Thomas Jefferson National Accelerator Facility, Newport News, Virginia 23606, USA}
\author{H. Kamano}
\affiliation{Research Center for Nuclear Physics, Osaka University, Osaka, Japan}
\author{T.-S. H. Lee}
\affiliation{Physics Division, Argonne National Laboratory, Argonne, Illinois 60439, USA}
\author{T. Sato}
\affiliation{Department of Physics, Osaka University, Toyonaka, Osaka 560-0043, Japan}
\affiliation{J-PARC Branch, KEK Theory Center, Institute of Particle and Nuclear Studies,
High Energy Accelerator Research Organization (KEK), Tokai, Ibaraki 319-1106, Japan}

\date{\today}

\begin{abstract}
We have investigated the model dependence of  meson resonance properties
extracted from the Dalitz-plot analysis of
the three-pions photoproduction reactions on the nucleon. Within a unitary 
model developed in our earlier work, we generate 
Dalitz-plot distributions as data to perform an isobar model fit that is similar to most of
the previous analyses of three-pion production reactions. It is found that
the resonance positions from the two models agree well when both fit
the data accurately, 
except for the resonance poles near branch points.
The residues of the resonant amplitudes
extracted from the two models and by the usual Breit-Wigner procedure
 agree well only for the isolated resonances with narrow
widths. For overlapping resonances, most of the extracted residues could be drastically different.
Our results suggest that even with high precision data, the resonance extraction
should be based on models within which the amplitude parametrization is
constrained by three-particle unitarity condition.
\end{abstract}



\maketitle

\section{\label{sec:intro} Introduction}
Meson properties are important
information for understanding
the confinement mechanism of QCD. Thus the investigation of
meson spectroscopy has long been an important subject in hadron physics.
In recent years,  more emphasis has been placed on the study of
 mesons with quantum numbers beyond the classification of conventional
constituent quark model. Such mesons, called exotic mesons, are expected
to have  explicit gluonic and/or four-quark components in their
structure~\cite{klempt}. Therefore the search for exotic mesons
has been an important goal 
in the experiments on $\pi N\to M^* N\to \pi\pi\pi N$
at BNL~\cite{e852} and CERN~\cite{compass}, and 
$\gamma N\to M^* N\to \pi\pi\pi N$ at JLab~\cite{clas}, where the
intermediate excited mesons $M^*$ could be exotic.
However, the existence  of  exotic mesons,
such as  $\pi_1(1600)$ ($J^{PC}=1^{-+}$)
 has not been conclusive so far.
The forthcoming experiments to be performed at JLab after the 12~GeV
upgrade~\cite{gluex} are aimed at providing high precision  data
for making progress in this direction.
In addition to searching for exotic mesons, the new data
can also be useful for investigating some mesons that
could have exotic structure~\cite{3p0},
and can 
be revealed in their characteristic decay patterns, as
discussed in Ref.~\cite{3p0} with the $^3P_0$ model.

\begin{figure}[t]
\includegraphics[width=\textwidth]{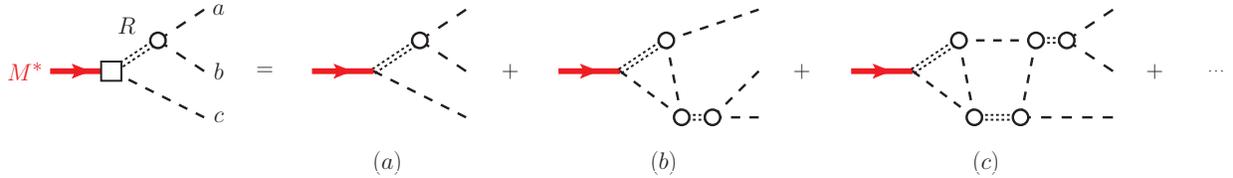}
\caption{\label{fig:mstar-decay3}
(color online) $M^*$-decay amplitude.
}
\end{figure}

We are here interested in the excited mesons that decay 
into three light mesons ($\pi\pi\pi$, $\pi\pi K$, etc.). 
Since these excited mesons are unstable and couple with multi-mesons
continuum to form resonances, the meson spectroscopy can be determined
only by analyzing the resonances extracted from the
meson production reaction data.
Conventionally,  these data were analyzed by using
the so-called isobar model (IM)
in which two of the three mesons form a light flavor excited meson
$R$ ($f_0$, $\rho$, $K^*$, etc.) and the third meson is treated as a
spectator of the propagation and the decay of $R$ into two light mesons.
This approach obviously violates the three-body unitarity and  neglects
the coupled-channels effects since the outgoing $R$ can have multiple scattering
with the third meson,
as illustrated in Fig.~\ref{fig:mstar-decay3}.

Recently we developed a unitary coupled-channels model~\cite{3pi}.
In this work, we use our model to analyze
the $\gamma p \rightarrow M^* n\rightarrow \pi^+\pi^+\pi^- n$ reaction, 
and study the importance of three-body unitarity and coupled-channels
effects in extracting the excited meson properties.
We consider the reaction where the intermediate $M^*$ can
be several and can overlap.
We will show that while the IM can fit the same Dalitz plot data
generated from our unitary model (UM), 
the extracted resonance parameters are 
rather different. Our finding indicates the limitation of
the IM in establishing the meson spectroscopy.

This paper is organized as follows.
In Sec.~\ref{sec:formulation}, we present formulae based on our model
for identifying the meson resonances in the amplitudes of
the $\gamma p\to M^*n\to \pi^+\pi^+\pi^-$ reaction, and give the expression
for calculating the corresponding
Dalitz plots of the cross sections.
Our procedure and numerical results are presented in 
Sec.~\ref{sec:result}, followed by a summary in 
Sec.~\ref{sec:summary}.

\section{\label{sec:formulation} Formulation}

\begin{figure}
\includegraphics[width=150mm]{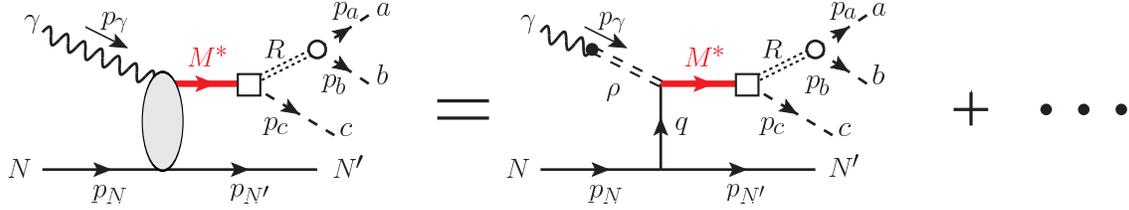}
\caption{\label{fig:gp3pi} (color online)
A graphical representation of the
$\gamma N\to abc N^\prime$ reaction. The momentum for each particle is
 shown to explain the notation used in this work.
The white square indicates the dressed vertex more explicitly shown in
 Figs.~\ref{fig:decay-vertex} and \ref{fig:eq}.
}
\end{figure}

In this paper, we denote pions by 
$a$, $b$, $c$ and the light excited mesons (such as $f_0, \rho $  ) by
$R$.  As illustrated in the left-hand-side of 
 Fig.~\ref{fig:gp3pi}, 
we assume that the $\gamma N\to abc N^\prime$  reaction proceeds via the photo-production of 
resonant $M^*$  states which
decay into $c R$ states. Under the three-particle unitarity condition,
the propagating $c R$ states experience the multiple-scattering due to the
$Z$-diagram mechanism, as illustrated in Fig.~\ref{fig:mstar-decay3}.
The final three pion states are then generated via 
the  $R \rightarrow \pi\pi$ decay
mechanism. 
In the vector meson dominance (VDM) model, one of the
production mechanisms can be calculated from a $\pi\rho\rightarrow M^*$ 
and the well known $\pi NN$ vertex, as  illustrated in the right-hand-side of 
Fig.~\ref{fig:gp3pi}.
For simplicity, we will calculate the cross section of 
$\gamma N\to abc N^\prime$ reaction using this VDM mechanism, 
and the $\pi\rho\rightarrow M^*$ vertex
that reproduce partial width
predicted by the $^3P_0$ model
of Ref.~\cite{3p0}. 
Accordingly all $M^*\rightarrow \pi R$ couplings included in
our calculations are also fixed in the same manner.
Obviously, this is a simplification, but is sufficient
for our present purposes in investigating the importance of
three-particle unitarity in extracting $M^*$ from the
three-pion distribution data. For analyzing the data from CLAS, we
need to develop models of other mechanisms. This non-trivial task
is beyond the scope of this investigation.

\subsection{\label{sec:xsc} Cross section}
\label{subsec:cs}

With the momentum variables specified in Fig.~\ref{fig:gp3pi}(b),
the cross section of $\gamma N \rightarrow \pi\pi\pi N$
can be written as
\begin{eqnarray}
d\sigma&=&
{1\over v_{\text{rel}}}
{1\over 2 E_\gamma}
{1\over 2 E_N}
(2 m_N)^2
{{\cal B}\over 4}
\sum_{s_N,s_{N^\prime},\lambda_\gamma}
|{\cal M}_{\gamma N\to \pi\pi\pi N^\prime}|^2 \nonumber \\
&\times&
(2\pi)^4 \delta^{(4)}(p_i-p_f) {d^3p_{N^\prime}\over (2\pi)^3 2 E_{N^\prime}}
{d^3p_a\over (2\pi)^3 2 E_a}
{d^3p_b\over (2\pi)^3 2 E_b}
{d^3p_c\over (2\pi)^3 2 E_c} \label{eq:crst-3pi} \ ,
\end{eqnarray}
where $v_{\text{rel}}$ is the relative velocity of the initial particles.
We used and will use a notation $E_x=\sqrt{m_x^2 + \vec{p}^2_x}$
to denote the free energy for a particle $x$ with the mass $m_x$ and the
momentum  $\vec{p}_x$.
We denote the Bose factor for the final three pions by
${\cal B}$, and
${\cal B} = 1/2$ for $\pi^+\pi^+\pi^-$ final states.
We have taken the average of the initial nucleon and photon
polarizations
(${1\over 4}\sum_{s_N,\lambda_\gamma}$) and the summation over the final nucleon spin
($\sum_{s_{N^\prime}}$).

We perform the calculations in the center of mass (CM) frame of the total 
system. 
The orientation of the three final pions are specified by the Euler angles
$\Omega_{\rm Euler}=(\alpha,\beta,\gamma)$ in the three-meson CM  frame
(see Eq.~(10) and Fig.~1 of Ref.~\cite{kuhn} for the definition of the Euler angles).
With some manipulation of the four-body phase-space 
factor~\cite{chung,kuhn}, Eq.~(\ref{eq:crst-3pi}) leads to
\begin{eqnarray}
\label{eq:dalitz-unpol}
{d\sigma \over dt\ dW\ d\Omega_{\rm Euler}\ d m^2_{ab}\ d m^2_{bc}}
&=&{{\cal B} m_N^2\over 4 \sqrt{ (p_\gamma\cdot p_N)^2}}
{1\over (4\pi)^7} {1\over E_{\text{tot}}\ p^{\text{cm}}_N W}
\sum_{s_N,s_{N^\prime},\lambda_\gamma}
|{\cal M}_{\gamma N\to \pi\pi\pi N^\prime}|^2 \ ,
\label{eq:dalitz-unp}
\end{eqnarray}
where $d \Omega_{\rm Euler} = d\alpha\ d \cos\beta\ d\gamma$, 
$E_{\text{tot}}$ is the total energy, $p^{\text{cm}}_N$ is  the initial nucleon
momentum,  $t=(p_N-p_{N^\prime})^2$,
$W=\sqrt{(p_a+p_b+p_c)^2}$, and
$m^2_{ij}=(p_i+p_j)^2$.
For a given set of $(t,W,\Omega_{\rm Euler})$, 
we then can calculate the Dalitz-plot distributions of the three out-going pions
as functions of two-particle invariant masses $m_{ab}$ and $m_{bc}$
 by using Eq.~(\ref{eq:dalitz-unp}).

\begin{figure}[t]
\includegraphics[width=150mm]{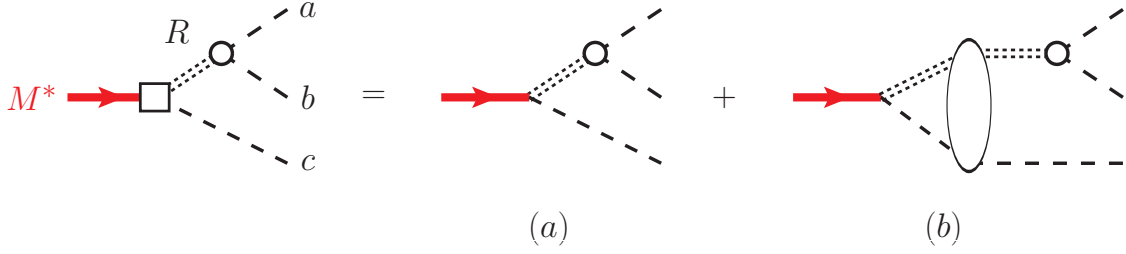}
\caption{\label{fig:decay-vertex}
(color online)
A graphical representation of the
dressed $M^*\rightarrow cR$ vertex}
\end{figure}

\begin{figure}[t]
\includegraphics[width=150mm]{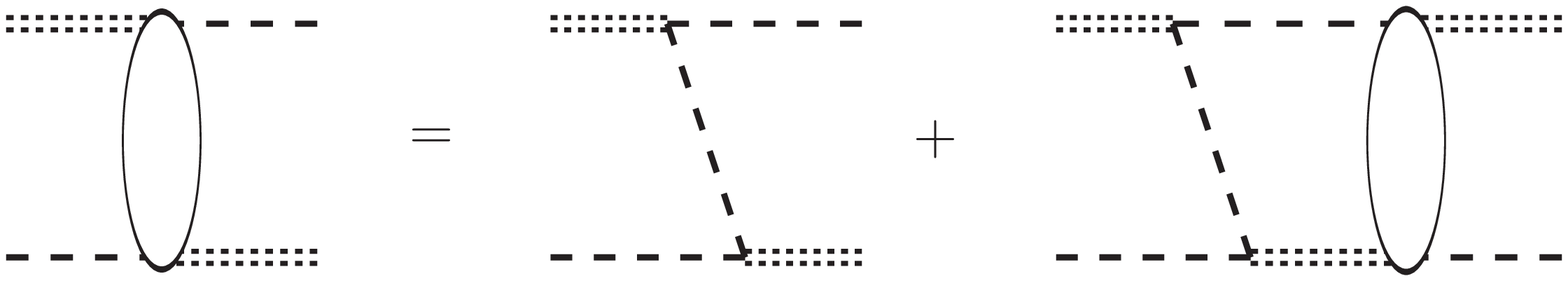}
\caption{\label{fig:eq}
A graphical representation of $cR$ scattering equation.}
\end{figure}

\subsection{\label{sec:me} Matrix element}
\label{subsec:umodel}
With the normalization $\inp{\vec{k}}{\vec{k}^{\,\,'}}=\delta(\vec{k}-\vec{k}^{\,\,'})$
for the  plane-wave state,
the invariant amplitude 
${\cal M}_{\gamma N\to \pi\pi\pi N^\prime}$ in Eq.~(\ref{eq:dalitz-unp})
is related to the scattering amplitude $T_{abc N^\prime,\gamma N}$  by
\begin{eqnarray}
{\cal M}_{\gamma N\to abc N^\prime}
&=& - {1\over (2\pi)^3}
\sqrt{(2\pi)^3 2E_\gamma}\sqrt{(2\pi)^3 2E_N} \nonumber \\
&\times&
 \sqrt{(2\pi)^3 2E_a(p_a)} \sqrt{(2\pi)^3 2E_b(p_b)} \sqrt{(2\pi)^3 2E_c(p_c)}
 \sqrt{(2\pi)^3 2E_{N^\prime} (p_{N^\prime})} \
T_{abc N^\prime,\gamma N} \ .
\label{eq:stmx}
\end{eqnarray}
With the VDM process illustrated in the right-hand-side of 
 Fig.~\ref{fig:gp3pi},
the scattering amplitude  in Eq.~(\ref{eq:stmx})
can be written within our model as
\begin{eqnarray}
\label{eq:t-matrix}
T_{abc N^\prime,\gamma N} &=& 
\sum_{\alpha}
\sum_{i_\alpha j_\alpha}\sum_{S^z_{M^*}}
\sum^{\text{cyclic}}_{(a'b'c')} F_{(a'b')c',M^\ast_{i_\alpha}}(W)\
[G_{M^*}(W)]_{i_\alpha j_\alpha}\ T_{M^*_{j_\alpha} N',\gamma N}(W)
\ , \label{eq:gn-3pin}
\end{eqnarray}
where $T_{M^* N',\gamma N}$ is the $M^*$ photo-production amplitude,
$F_{(ab)c,M^\ast}(W)$ is the $M^*\rightarrow abc$ decay amplitude, and
 $G_{M^*}(W)$ is the propagator of the $M^*$ resonant state.
The index $\alpha$ labels a set of quantum numbers (spin,
isospin, parity) of $M^*$, and 
$i_\alpha, j_\alpha$ runs over all $M^*$s belonging to the set of
quantum numbers $\alpha$.
The summation over the $M^*$ spin orientation is denoted by $\sum_{S^z_{M^*}}$.
The symbol $\sum^{\text{cyclic}}_{(a'b'c')}$
means taking summation over the cyclic permutation, 
$(a'b'c') = (abc), (cab), (bca)$.
As illustrated in Fig.~\ref{fig:decay-vertex}, the $M^*$ decay amplitude in Eq.~(\ref{eq:t-matrix}) 
consists of two terms
\begin{eqnarray}
F_{(ab)c,M^*}(E) = F^{\text{DIR}}_{(ab)c,M^*}(E) 
+ F^{\text{FSI}}_{(ab)c,M^*}(E)\,.
\label{eq:decay-amp0}
\end{eqnarray}
The direct decay amplitude [Fig.~\ref{fig:decay-vertex}(a)] is defined by
\begin{eqnarray}
F^{\text{DIR}}_{(ab)c,M^*}(E) &=&
\sum_{R} \sum_{c'R'} \bra{ab}f_{ab,R} G_{cR,c'R'}(E) \Gamma_{c'R',M^*}\ket{M^*} ,
\label{eq:t-isobar}
\end{eqnarray}
where $f_{ab,R}$ is the  $R\rightarrow ab$ vertex interaction, and
$\Gamma_{c'R',M^*}\ket{M^*}$ describes the $M^*\rightarrow c'R'$ decay.    
The summation in Eq.~(\ref{eq:t-isobar}) runs over the particle species and its momentum,
spin, and isospin components.
The second term of Eq.~(\ref{eq:decay-amp0}) [Fig.~\ref{fig:decay-vertex}(b)]
 includes the final state    
interaction (FSI),  as required by the 
 three-particle unitarity condition. It has the following expression:
\begin{eqnarray}
F^{\text{FSI}}_{(ab)c,M^*}(E) &=&
\sum_{R}\sum_{c'R'} \sum_{c'''R''',c''R''} \bra{ab} f_{ab,R} G_{c R,c'R'}(E)
T'_{c'R', c'''R'''}(E) G_{c'''R''',c''R''}(E) \Gamma_{c''R'',M^*}
\ket{M^*} ,
\nonumber\\
\label{eq:decay-int}
\end{eqnarray}
where $T'_{c'R', cR}(E)$ is the
$cR \rightarrow c'R'$ scattering amplitudes. As illustrated in
 Fig.~\ref{fig:eq},  $T'_{c'R', cR}(E)$ is defined by 
the following coupled-channels scattering equation,
\begin{equation}
T'_{c'R',cR}(E) = Z_{c'R',cR}(E) +
\sum_{c'''R''',c''R''}  Z_{c'R',c'''R'''}(E) 
G_{c'''R''',c''R''}(E) T'_{c''R'',cR}(E)\,, 
\label{eq:rpi}
\end{equation}
where $Z_{c'R',cR}$ is the one-particle-exchange
$Z$-diagram interaction
that is  also determined by the $R\to ab$ vertex interaction
$ f_{ab,R} $ 
\begin{equation}
Z_{c'R',cR}(E) = \sum_{c''}
f_{R', cc''}  \frac{1}{E - E_c - E_{c'} - E_{c''} +i\epsilon}
f_{c'c'',R} \ ,
\label{eq:z}
\end{equation}
where $c''$ is the exchanged meson.

The Green function in Eqs.~(\ref{eq:t-isobar})-(\ref{eq:rpi}) is defined by
\begin{equation}
\left[G^{-1}(E)\right]_{c'R',cR} =
\delta_{c',c}\left[ (E - E_c - E_R)\delta_{R',R} - \Sigma_{R',R}(E - E_c) \right] \ ,
\label{eq:pir-G0}
\end{equation}
where the self-energy of light excited-meson $R$ is determined by the
$R\rightarrow ab$  vertex interaction
\begin{equation}
\Sigma_{R',R}(w) =
\sum_{ab=\pi\pi,K\bar K}
\bra{R'} f_{R',ab}
 \frac{{\cal B}_{ab}}{w- E_{a}- E_{b}+i\epsilon} f_{ab,R}\ket{R} .
\label{eq:r-selfe}
\end{equation}
Here ${\cal B}_{ab}$ is a factor associated with the Bose symmetry of mesons:
${\cal B}_{ab} = 1/2$ if $a$ and $b$ are the identical particles or otherwise ${\cal B}_{ab}= 1$.

The $M^*$ propagator in Eq.~(\ref{eq:t-matrix}) is defined by
  \begin{eqnarray}
\left[G^{-1}_{M^*}(E)\right]_{ij} = (E- M^0_{M^*_i})\delta_{ij} - \left[\Sigma_{M^*}(E)\right]_{ij}\,,
\label{eq:mstar-g1}
\end{eqnarray}
where $ M^0_{M^*_i}$ is a bare mass, and the self energy is given by
\begin{eqnarray}
[\Sigma_{M^*}(E)]_{ij} &=& \sum_{cR,c'R'} \bra{M^*_i} \Gamma_{M^*_i,cR}G_{c R,c'R'}(E)
\ket{\bar{\Gamma}_{c'R',M^*_j}}.
\label{eq:mstar-sigma}
\end{eqnarray}
The self energy $[\Sigma_{M^*}(E)]_{ij}$ is non-vanishing only when $M^*_i$
and $M^*_j$ have the same spin, isospin and parity.
In Eq.~(\ref{eq:mstar-sigma}), $\ket{\bar{\Gamma}_{cR,M^*}}$ is the
 dressed $M^*\to c R$ vertex function defined by
\begin{eqnarray}
\ket{\bar{\Gamma}_{cR,M^*}}
= \sum_{c'R'} \left[\delta_{cR,c'R'} + \sum_{c''R''}T'_{cR,c''R''} G_{c''R'',c'R'}(E)\right]
\Gamma_{c'R',M^*}\ket{M^*}.
\label{eq:mf-dressed}
\end{eqnarray}
Note that the $M^*\rightarrow abc$ amplitude $F_{(ab)c,M^*}$
defined by Eqs.~(\ref{eq:decay-amp0})-(\ref{eq:decay-int}) is identical to the matrix element
$\sum_{Rc'R'}\bra{ab}f_{ab,R}G_{cR,c'R'}\ket{\bar{\Gamma}_{c'R',M^*}}$.

To proceed, we  need to define $R\leftrightarrow ab$ and
$M^*\rightarrow cR$ vertex functions.
The $R\rightarrow ab$ vertex
functions $f_{ab,R}$  have been determined within our model~\cite{3pi} by
fitting the $\pi\pi$
scattering amplitudes up to 2 GeV with $R = f_0, \rho, f_2$.
We include $\pi\pi$ and $K\bar{K}$ channels in our model of $\pi\pi$ scattering.
While the $\pi\pi\pi$ three-body states appear in $Z$-diagram interactions in Eq.~(\ref{eq:z}),
$\pi K\bar K$ states appear only in the Green function defined by Eq.~(\ref{eq:pir-G0}).

We thus will only consider $cR = \pi f_0, \pi \rho, \pi f_2$ channels.
For the bare $M^*\leftrightarrow c R$ interaction 
$\Gamma_{cR,M^*}(\vec{p}_c)$,
we use a partial wave expansion,
\begin{eqnarray}
\label{eq:vertex}
\Gamma_{cR,M^*}(\vec{p}_c) =
\sum_{l} \inp{t_c t^z_c t_R t^z_{R}}{T_{M^\ast} T^z_{M^\ast}}
\inp{ll^z, s_R s^z_{R}}{S_{M^\ast} S^z_{M^\ast} }
Y^*_{l,l^z}(-\hat{p}_c) \Gamma_{(cR)_l,M^\ast}(p_c) \ ,
\end{eqnarray}
where the first (second) parenthesis is isospin (angular momentum)
Clebsch-Gordan coefficient;
$s_x$ ($t_x$) is the spin (isospin) of a particle $x$ and 
$s^z_x$ ($t^z_x$) is its $z$-component, and $l$
is the orbital angular momentum of the relative $c R$ motion.
The vertex function $\Gamma_{(cR)_l,M^\ast}(p)$ is given with 
the following parametrization for each partial-wave:
\begin{eqnarray}
\Gamma_{(cR)_l,M^*}(p) =
\frac{1}{(2\pi)^{3/2}} C_{(cR)_l,M^*}
\sqrt{m_R\over 4 E_R(p) E_\pi(p)}
\left(
\frac{\Lambda^2_{(cR)_l ,M^*}}{p^2 + \Lambda^2_{(cR)_l,M^\ast}}
\right)^{2+(l/2)}
\left(\frac{p}{m_\pi}\right)^{l} \ .
\label{eq:bare_mstar}
\end{eqnarray}
The parameters $C_{(cR)_l,M^*}$ and $\Lambda_{(cR)_l,M^\ast}$
are chosen to reproduce the partial decay widths of $M^*\rightarrow \pi R$
predicted by the $^3P_0$ model of Ref.~\cite{3p0}, and will be
explained in the next section.

To calculate the $\gamma N \rightarrow \pi\pi\pi N'$ amplitude
defined by  Eq.~(\ref{eq:gn-3pin}), we also need to calculate 
the VDM photo-production amplitude 
$T_{M^* N',\gamma N}$ illustrated in the right-hand-side of Fig.~\ref{fig:gp3pi}. 
We use the following interaction Lagrangian to describe the emission of
the pion from the nucleon:
\begin{eqnarray}
{\cal L}_{\pi NN} = - {f_{\pi NN}\over m_\pi} \bar \psi_N
 \gamma_\mu\gamma_5 \vec\tau\cdot \psi_N \partial^\mu \vec\pi \ ,
\label{eq:lag-pinn}
\end{eqnarray}
where $f^2_{\pi NN}(q^2=0)/4\pi=0.08$ and 
$f_{\pi NN}(q^2)/f_{\pi NN}(0)=1/(1-q^2/\Lambda^2_{\pi NN})^2$.
We use $\Lambda_{\pi NN} = 700$~MeV which is close to most of
the values from the $\pi N$ scattering model~\cite{sl96}.
The photon-$\rho$ contact interaction within the VDM model is defined
by the following Lagrangian:
\begin{eqnarray}
{\cal L}_{\rho\gamma} =  {e m^2_\rho \over g_\rho} \rho^3_\mu A^\mu \ ,
\label{eq:lag-vdm}
\end{eqnarray}
with $g^2_\rho/(4\pi)=2.2$.

With the Lagrangians Eqs.~(\ref{eq:lag-pinn}) and~(\ref{eq:lag-vdm}),
the $\gamma N \rightarrow M^* N'$ production amplitude in $\gamma N$ CM
frame is of the following form
\begin{eqnarray}
\label{eq:production}
T_{M^* N',\gamma N}(\vec{p}_{M^*}\vec{p}_{N'}; \vec{p}_\gamma,\vec{p}_N) &=&
\sqrt{1\over 2 E_\gamma E_N(\vec{p}_N) E_{N}(\vec{p}_{N'})}
\sum_i \sqrt{4 E_{\rho_i}(q) E_\pi(q)}
\Gamma_{M^*,\pi\rho_i} (\vec q)
 \left(e\over g_\rho\right) \nonumber \\
&& \times {1\over q^2-m^2_\pi}
 {i f_{\pi NN}(q^2)\over m_\pi}
\braketa{\tau_x}\
\bar u_{\vec{p}_{N'}} \Slash{q}\gamma_5 u_{\vec{p}_N} \ ,
\label{eq:opep}
\end{eqnarray}
where $q = p_N - p_{N'}$, $u_{\vec{p}}$ is the nucleon spinor. 
The summation is taken over the first and second bare $\rho$ states.
The isospin matrix element
between the nucleon states is denoted by
$\braketa{\tau_x}=1\ (\sqrt{2})$ for $N=N'=p$ ($N=p,N'=n$).
The $M^*\rightarrow \rho_i\pi$ vertex function 
$\Gamma_{M^*,\pi\rho_i} (\vec q)$ has been defined by 
Eqs.~(\ref{eq:vertex}) and (\ref{eq:bare_mstar}).

\subsection{Isobar-model}
\label{subsec:isobar}
We can obtain a model similar to 
the commonly used IM from the above
formula by neglecting the $\pi R$ final state interactions, $F^{\text{FSI}}_{(ab)c,M^*}$ in
Eq.~(\ref{eq:decay-amp0}), which are due to the $Z$-diagram
mechanism, as illustrated in Fig.~\ref{fig:decay-vertex}(b) and Fig.~\ref{fig:eq}.
Furthermore, the $M^*$ propagator $G_{M^*}(W)$ in Eq.~(\ref{eq:gn-3pin})
are replaced by a Breit-Wigner (BW) form.
Explicitly we consider the following expression of the IM,
\begin{eqnarray}
\label{eq:t-matrix-bw}
T^{\text{IM}}_{abc N^\prime,\gamma N} &=&
\sum_{\alpha}
\sum_{j_\alpha}\sum_{S^z_{M^*}}
\sum^{\text{cyclic}}_{(a'b'c')} F^{\text{DIR}}_{(a'b')c',M^\ast_{j_\alpha}}(W)\
[G^{\rm BW}_{M^*}(W)]_{j_\alpha j_\alpha}\ T_{M^*_{j_\alpha} N',\gamma N}(W)
\ , \label{eq:gn-3pin-im}
\end{eqnarray}
with
\begin{eqnarray}
\label{eq:BW}
[G^{\rm BW}_{M^*}(W)]_{jj} = -{M^j_{\text{BW}}\over (M^{j}_{\text{BW}})^2 - W^2 - i
M^j_{\text{BW}}\Gamma^j_{\text{BW}}
\left[\left({q\over q^j_{\text{BW}}}\right)^{2(L^j+1)}
\left({(q^{j}_{\text{BW}})^2+(\Lambda^j_{\text{BW}})^2\over q^2+(\Lambda^j_{\text{BW}})^2}\right)^2
\right] }
\ ,
\end{eqnarray}
where $q$ is the on-shell momentum satisfying the equation:
\begin{eqnarray}
W = \sqrt{m_\pi^2 + q^2} + \sqrt{m_{R_{\text{BW}}}^2 + q^2} \ ,
\end{eqnarray}
and $m_{R_{\text{BW}}} = 770$~MeV; $q=q^j_{\text{BW}}$ when $W=M^j_{\text{BW}}$.
The integer $L^j$ is the lowest allowed orbital angular momentum between
the pion and a vector boson for a given $M^*_j$ at rest.
It is noted that the BW $M^*$ propagator in Eq.~(\ref{eq:BW})
is diagonal with respect to the index $j$, while 
the $M^*$ propagator for the UM given in
Eq.~(\ref{eq:mstar-g1}) can have off-diagonal component, connecting
different $M^*$s belonging to the same quantum number.
Furthermore, whereas the BW $M^*$ propagator in Eq.~(\ref{eq:BW}) is
purely phenomenological, 
the $M^*$ propagator for the UM needs to be
given by Eq.~(\ref{eq:mstar-g1}) as a consequence of three-body
unitarity.
The IM defined by Eqs.~(\ref{eq:t-matrix-bw}) and~(\ref{eq:BW})
will be used to extract the resonance
parameters from fitting the ``data'' generated from our UM
defined in Sec.~\ref{subsec:umodel}.

\section{Results}
\label{sec:result}

Our objective in this paper is to
 investigate the importance of the three-particle unitarity
in determining the excited meson $M^*$ properties from fitting
the Dalitz-plot distribution data of $3\pi$ states
for the $\gamma p\to M^* n\to \pi^+\pi^+\pi^- n$ reaction. 
We do not make an attempt to analyze the CLAS data~\cite{clas}
for this reaction here. 
Thus it is sufficient to consider
the VDM production mechanism illustrated in Fig.~\ref{fig:gp3pi}(b).
Then we can set up a model by fixing the parameters associated with
$M^*$ states using partial widths predicted by the $^3P_0$ model of Ref.~\cite{3p0}.
This will be explained in Sec.~\ref{sec:setup}.

Once the $M^*$ parameters are fixed, we can
perform the calculations using the formula presented in previous sections
since all parameters needed to calculate the 
$\pi R\rightarrow \pi R'$ amplitudes [Eq.~(\ref{eq:rpi})] and
$\pi R$ propagator [Eq.~(\ref{eq:pir-G0})]
 have been determined in our previous work~\cite{3pi}.
The Dalitz plots for the $\pi^+\pi^+\pi^- $ 
are calculated
using Eq.~(\ref{eq:dalitz-unp}), and are presented in Sec.~\ref{subsec:cal-dalitz}
In Sec.~\ref{subsec:fit}, we describe how the
generated Dalitz plots are used as the data to determine the parameters of the IM
described in Sec.~\ref{subsec:isobar}.

In Sec.~\ref{subsec:mstar}, we examine the differences between the resonance parameters
extracted with our UM and those from the fitted IM.
Their differences will indicate the importance of
the three-particle unitarity in determining the 
excited meson $M^*$ properties from fitting
the Dalitz-plot distribution data of the $3\pi$ states for the $\gamma p\to \pi^+\pi^+\pi^- n$
reaction.

\subsection{Determination of the $M^*$ parameters}
\label{sec:setup}

In our UM, we consider
the $\pi R$ partial waves that are found to be necessary to fit the
CLAS data for $1.0 < W < 1.36$~GeV~\cite{clas}.
We thus have  one or two bare $M^*$ states for four partial waves:
$J^{PC} = 1^{++}$ [$a_1(1230)$, $a_1(1700)$],
$2^{++}$ [$a_2(1320)$, $a_2(1700)$],
$2^{-+}$ [$\pi_2(1670)$, $\pi_2(1800)$],
$1^{-+}$ [$\pi_1(1600)$].
We assume that our bare $M^*$ states can be identified
with excited meson states of the $q\bar q$ excitation type
listed in Ref.~\cite{3p0},
and that their 
bare $M^*\to \pi R$ couplings are fixed so that
the partial decay widths predicted by the $^3P_0$ model~\cite{3p0} are
reproduced;
we use the formula given in Appendix I of Ref.~\cite{msl} to calculate
the partial widths.
We also assume that the daughter $\pi R$ states have the lowest allowed orbital
angular momenta.
The bare masses $M^0_{M^*}$ are also identified with the excited meson
masses listed in Ref.~\cite{3p0}.
The only exception is the $\pi_1(1600)$ that is speculated to be 
a hybrid state.
In this investigation we use the mass and 
partial widths for $\pi_1(1600)$ given in Ref.~\cite{IKP}.
For simplicity,
we set all $M^*\to \pi R$ cutoffs to $\Lambda_{(\pi R)_l, M^*} = 1$~GeV.
With the above specifications, the parameters for our UM are fixed.

\subsection{Calculations of Dalitz plots}
\label{subsec:cal-dalitz}

We use Eq.~(\ref{eq:dalitz-unp}) to generate 
the Dalitz-plot distributions of $\pi^+\pi^+\pi^-$ for 
the  $\gamma p \rightarrow \pi^+\pi^+\pi^- n$ reaction
at the photon energy $E_\gamma = 5$~GeV and the 
momentum-transfer $t = -0.4$~GeV$^2$. This is the kinematics considered in
the CLAS analysis~\cite{clas}.

We next  need to 
specify the Euler angles [$\Omega_{\rm Euler}$ 
in Eq.~(\ref{eq:dalitz-unpol})] that define
the orientation of the three-pion plane in their center of mass system.
Experimentally, it would be preferred to choose an orientation where the
three pions have less chance to interact with the final nucleon.
Thus we choose the Euler angle $\beta$ 
such that the three-pion plane 
is perpendicular to the direction of the final nucleon.
Because the cross section does not depend on $\alpha$, we set $\alpha=0$.
The remaining Euler angle $\gamma$ gives the rotation of
the three pions around their CM on the plane specified by $\alpha$ and $\beta$.
We calculate Dalitz plots by varying $\gamma$
in the range $0\le \gamma\le 2\pi$.

\begin{figure}
\includegraphics[width=100mm]{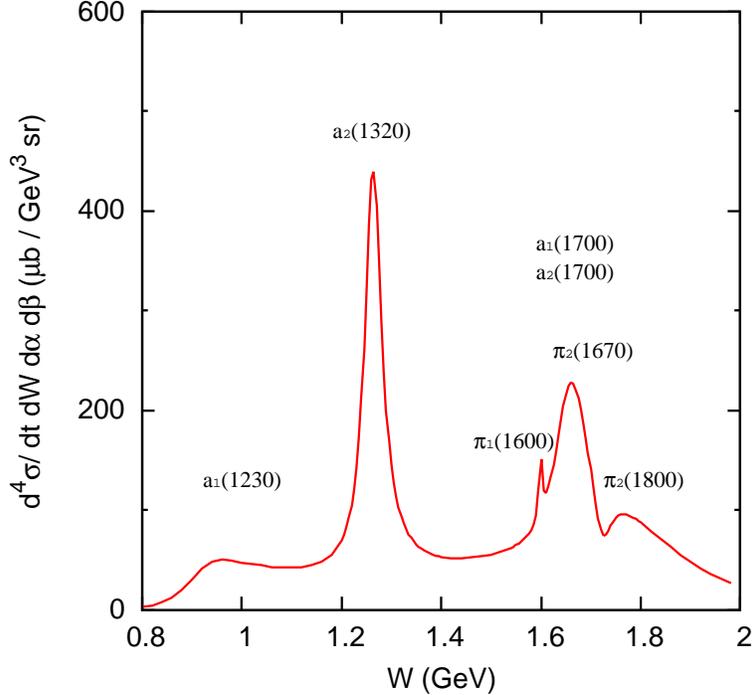}%
\caption{\label{fig:w}(color online) $W$-dependence of integrated 
Dalitz-plot distribution of the three pions for
the $\gamma p\to M^* n\to \pi^+\pi^+\pi^- n$ reaction,
obtained from Eq.~(\ref{eq:dalitz-unpol})
by integrating over
the kinematical variables $m^2_{ab}, m^2_{bc}$ and $\gamma$.
The other variables are
$t = -0.4$~GeV$^2$, $\alpha=0$,
and $\beta$ is fixed as explained in the text.
}
\end{figure}

To see the contributions of the considered resonances to  the generated
data, we show in Fig.~\ref{fig:w} the cross sections 
 calculated from Eq.~(\ref{eq:dalitz-unpol}) by
 integrating over $m^2_{ab}, m^2_{bc}$ and $\gamma$.
We  see a broad bump at $W\sim 0.95$~GeV due to $a_1(1230)$,
a highest peak at $W\sim 1.25$~GeV due to $a_2(1320)$.
The 
 second highest peak at $W\sim 1.65$~GeV is due to 
$a_1(1700)$ and $a_2(1700)$.
The gap at $W\sim 1.7$~GeV is due to an interference between 
$\pi_2(1670)$ and $\pi_2(1800)$.
The exotic $\pi_1(1600)$ is clearly visible as 
a small spike at $W\sim 1.6$~GeV.
Clearly it is a highly 
non-trivial task to fit the generated Dalitz-plot distribution data,
in particular, in the
region 1.6~GeV $\le W \le $ 1.8 GeV where the contributions from 
several resonances overlap strongly.

\begin{figure}
\begin{minipage}[t]{80mm}
\includegraphics[width=80mm]{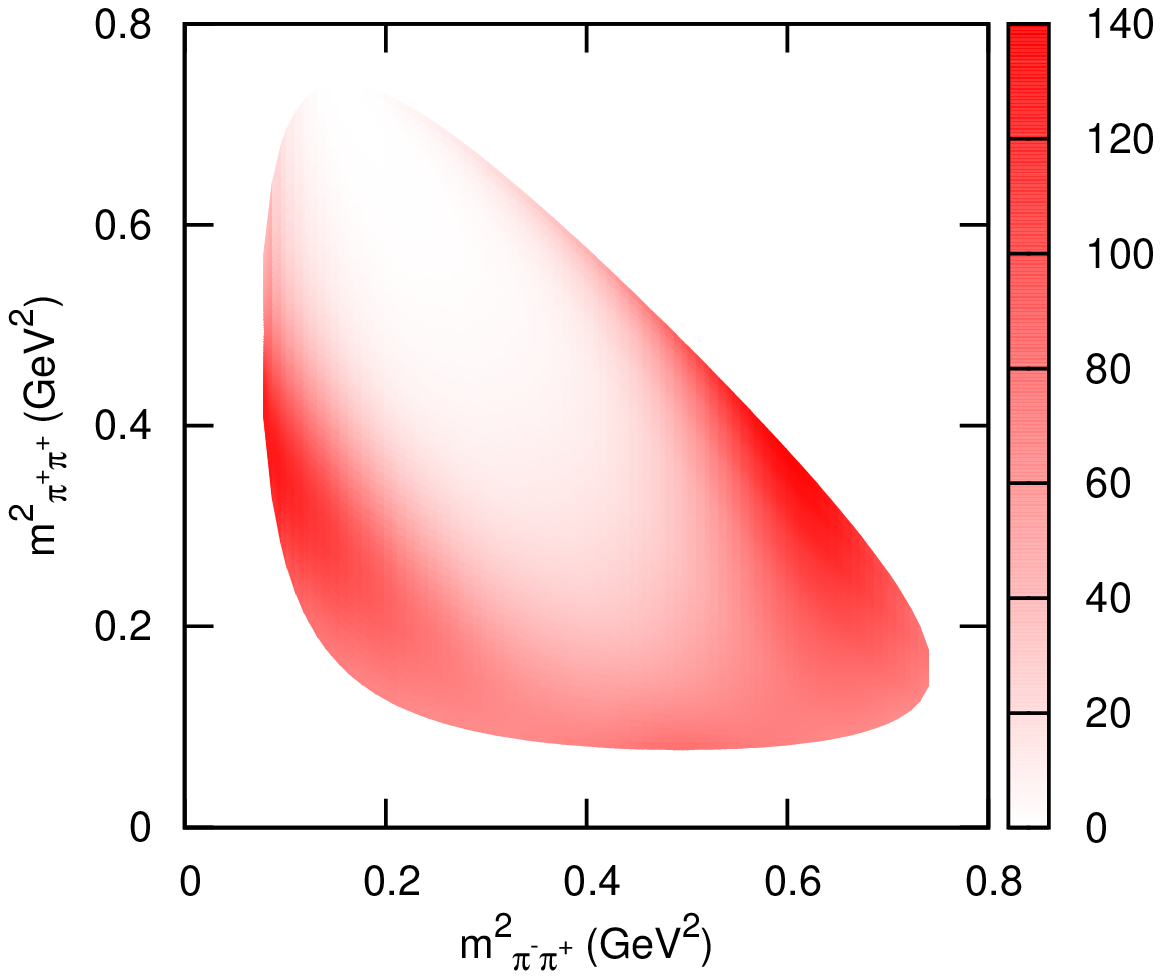}
\end{minipage}
\hspace{2mm}
\begin{minipage}[t]{80mm}
\includegraphics[width=77mm]{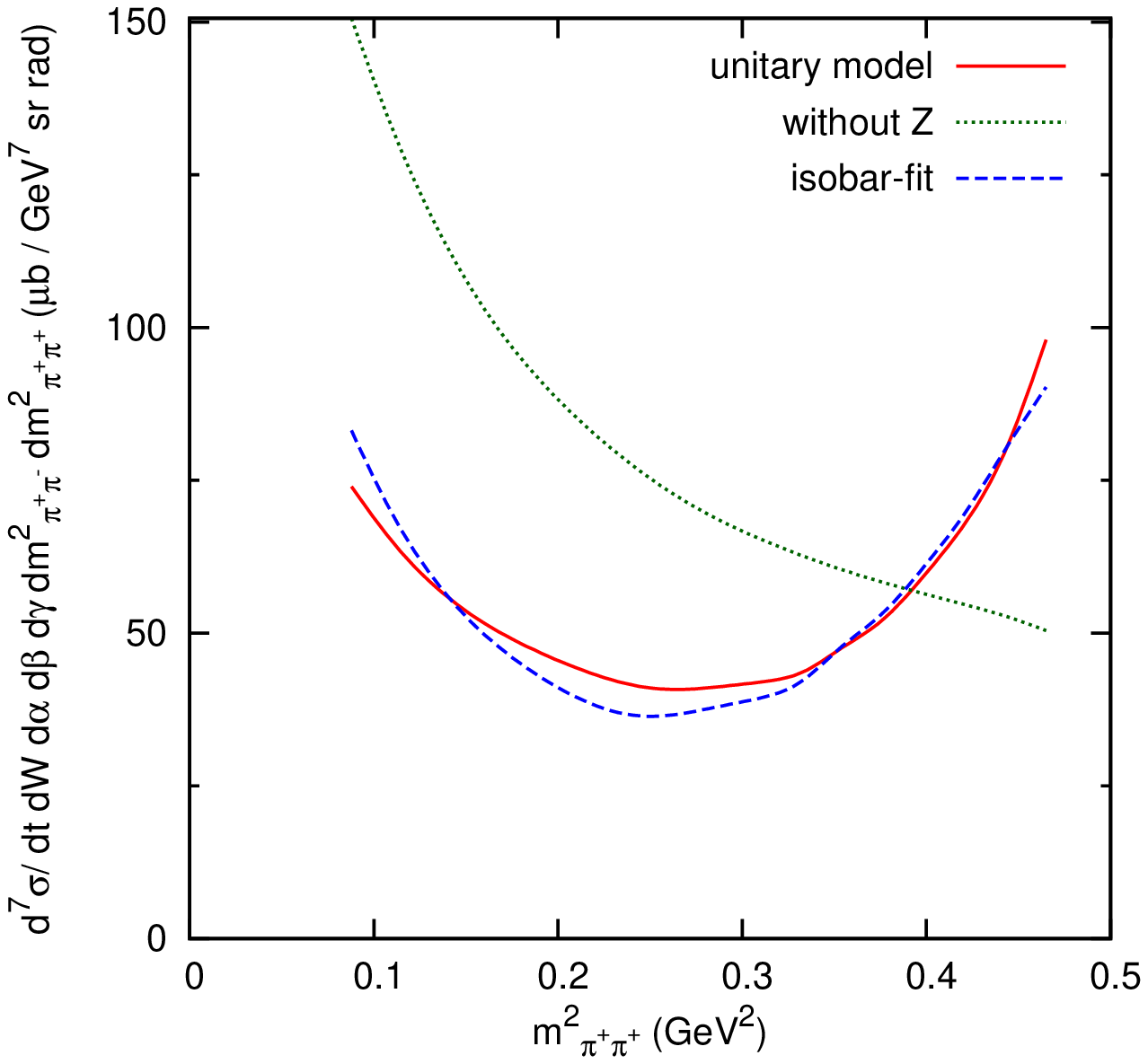}%
\end{minipage}
\caption{\label{fig:1000}
(color online)
(left) Dalitz-plot distribution of $3\pi$
[Eq.~(\ref{eq:dalitz-unpol}); unit: $\mu {\rm b}/ {\rm GeV}^7 {\rm sr\: rad}$]
from our UM for 
the $\gamma p\to \pi^+\pi^+\pi^- n$ reaction
at $W$ = 1~GeV near $a_1(1230)$ peak, $\cos\beta=-0.37$, and $\gamma=90^\circ$.
(right) One slice cut of the Dalitz plot on the left at 
$m^2_{\pi^-\pi^+} = 0.51$~GeV$^2$.
}
\end{figure}

\begin{figure}
\begin{minipage}[t]{80mm}
\includegraphics[width=80mm]{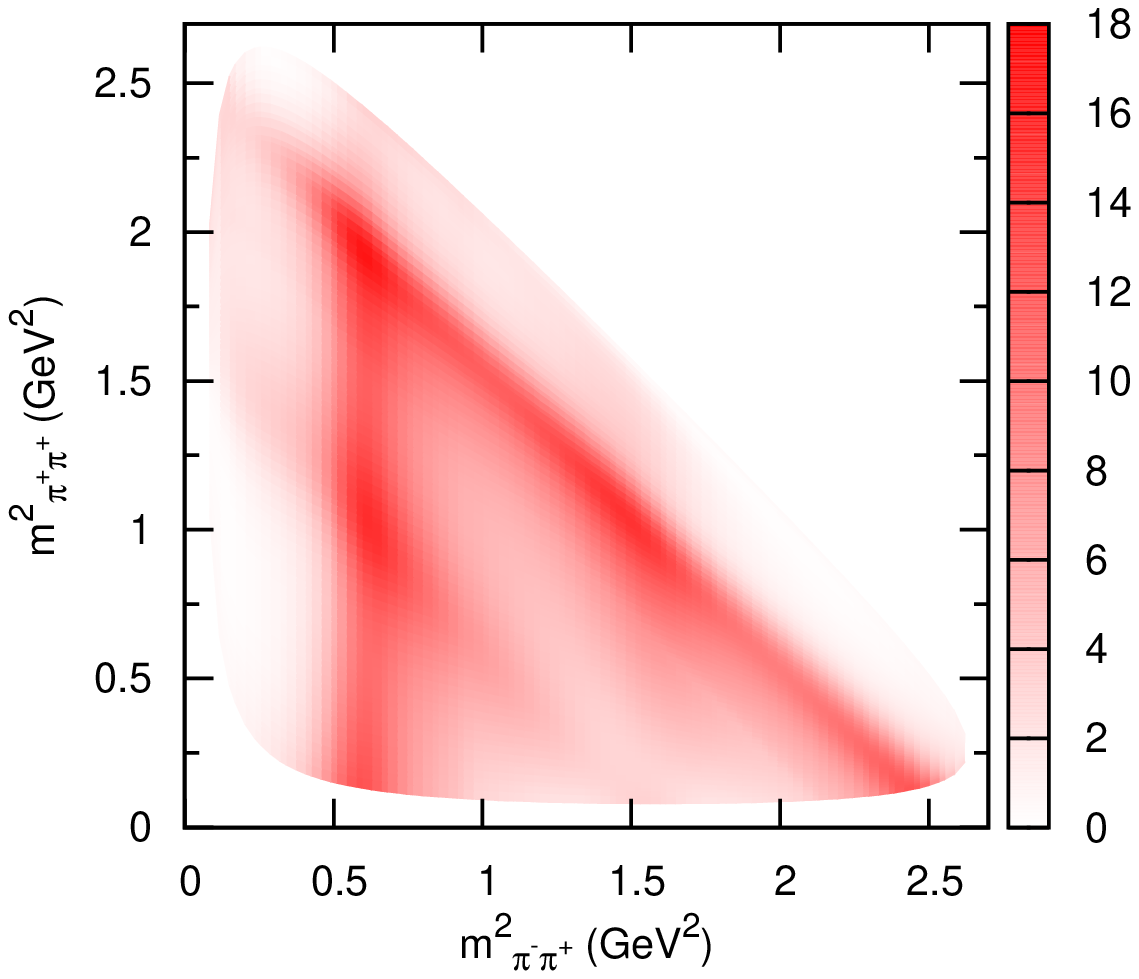}%
\end{minipage}
\hspace{2mm}
\begin{minipage}[t]{80mm}
\includegraphics[width=77mm]{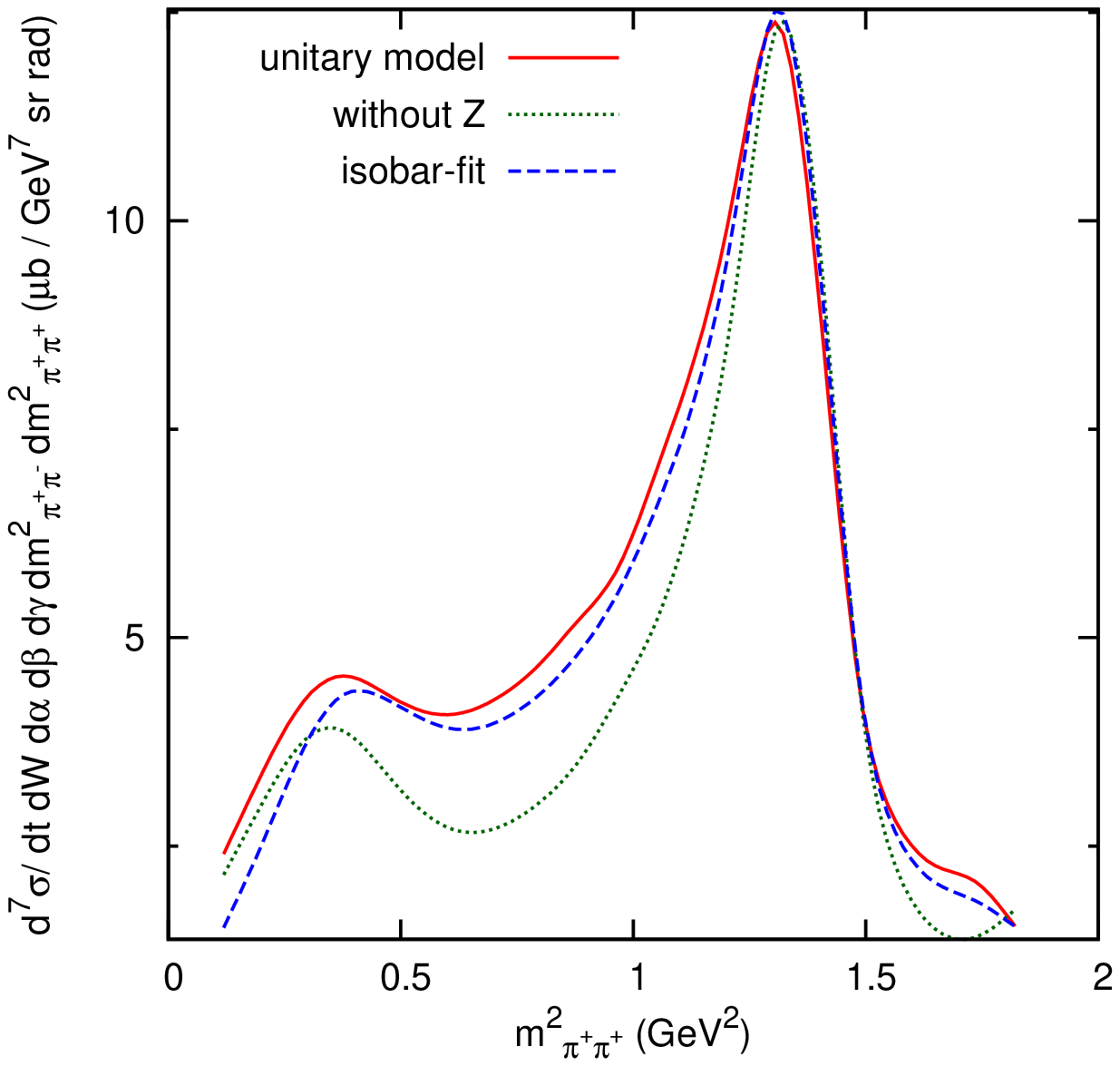}%
\end{minipage}
\caption{\label{fig:1760}
(color online)
Dalitz-plot distribution of $3\pi$ at $W$ = 1.76~GeV near $\pi_2(1800)$ peaks,
 $\cos\beta=-0.77$, and $\gamma=173^\circ$.
On the right, 
one slice cut of the Dalitz plot on the left is at 
$m^2_{\pi^-\pi^+} = 1.23$~GeV$^2$.
The other features are the same as those in Fig.~\ref{fig:1000}.
}
\end{figure}

\begin{table}[t]
\caption{\label{tab:pi2}
Bare masses $M^0_{M^\ast}$ (MeV)
as well as couplings $C_{(\pi R^{LI}_i)_l,M^*}$ (dimensionless) and
cutoffs $\Lambda_{\pi R,M^\ast}$ (MeV) of Eq.~(\ref{eq:bare_mstar}),
for the $M^\ast = \pi_2(1670)$, $\pi_2(1800)$, and $\pi_1(1600)$ from
the UM and those from the IM are compared. 
$R^{LI}_i$ denotes the $i$-th bare $R$ state
with the spin $L$ and the isospin $I$. $l$ denotes the orbital angular momentum
between $R^{LI}_i$ and $\pi$.
See Table I of Ref.~\cite{3pi} for the properties of $R^{LI}_i$.
For the IM,
the BW mass ($M_{\text{BW}}$), width ($\Gamma_{\text{BW}}$) and the cutoff
 ($\Lambda_{\text{BW}}$) are also listed.  }
\begin{ruledtabular}
\begin{tabular}{lcccccc}
  &\multicolumn{2}{c}{$\pi_2(1670)$} &\multicolumn{2}{c}{$\pi_2(1800)$} &\multicolumn{2}{c}{$\pi_1(1600)$} \\ \cline{2-7}
  &UM  & IM &UM  & IM &UM  & IM \\ \hline
$M^0_{M^*}$ ($M_{\text{BW}}$)          &1670.& 1815.		   &1800.&1727.            &1600.& 1599.	      \\          
       ($\Gamma_{\text{BW}}$)          & -   &  565.		   &   - &  69.		   &-    &    8.	      \\       	
($\Lambda_{\text{BW}}$)                & -   & 1005.		   &   - &1144.		   &-    &  651.	      \\ \hline	
$C_{(\pi R^{00}_1)_2,M^*}$      &   - &  0.18 +   0.05$i$  &0.13 &$-$ 0.08 + 0.02$i$&-    &              -\\       	
$\Lambda_{(\pi R^{00}_1)_2,M^*}$&   - & 1999.		   &1000.&1039.		   &-    &              -\\ \hline	
$C_{(\pi R^{00}_2)_2,M^*}$      &   - &  0.02 +   0.04$i$  &   - &$-$ 0.07 + 0.15$i$&-    &              -\\       	
$\Lambda_{(\pi R^{00}_2)_2,M^*}$&   - & 1991.		   &   - & 649.		   &-    &              -\\ \hline	
$C_{(\pi R^{11}_1)_1,M^*}$      &3.94 &  4.98 +   0.60$i$  &4.84 &$-$ 1.08 + 0.04$i$&1.01 &1.17 $-$ 0.60$i$ \\       	
$\Lambda_{(\pi R^{11}_1),M^*}$  &1000.& 1289.	           &1000.&1719.		   &1000 & 1116.	 \\\hline  	
$C_{(\pi R^{11}_2)_1,M^*}$      &   - & $-$ 2.29 $-$ 6.66$i$  &   - & 2.64 +   0.30$i$&   - & 0.07 + 0.16$i$ \\       	
$C_{(\pi R^{11}_2)_3,M^*}$      &   - & $-$ 0.02 $-$ 0.02$i$  &   - &$-$ 0.01 + 0.00$i$&-    &              -\\       	
$\Lambda_{(\pi R^{11}_2),M^*}$  &   - &  884.		   &   - &1007.		   &   - &  525.	      \\ \hline	
$C_{(\pi R^{20}_1)_0,M^*}$      &8.39 & 15.09 +   1.29$i$  &9.29 &$-$ 3.66 $-$ 0.55$i$&-    &-\\		       	
$C_{(\pi R^{20}_1)_2,M^*}$      &   - &  0.11 +   0.02$i$  &   - & 0.00 +   0.00$i$&-    &-\\		       	
$\Lambda_{(\pi R^{20}_1),M^*}$  &1000.&  817.		   &1000.&1077.            &-    &-
\end{tabular}
\end{ruledtabular}
\end{table}

\subsection{Fit}
\label{subsec:fit}

With the kinematics described above, the  Dalitz-plot distribution of 
the three pions are generated with Eq.~(\ref{eq:dalitz-unpol})
by varying  the variables 
$W$, $\gamma$ (one of the Euler angles), $m^2_{ab}$ and $m^2_{bc}$.
We calculate the cross sections at kinematical points that are uniformly
distributed in the space spanned by these kinematical variables. 
The calculated cross sections at these points are regarded as the `data' 
in determining the parameters of  the IM defined in Sec.~\ref{subsec:isobar}. 

As examples, we show in the left sides of
Figs.~\ref{fig:1000} and \ref{fig:1760} the
generated Dalitz-plot distributions
in the $m^2_{\pi^-\pi^+}$-$m^2_{\pi^+\pi^+}$ plane
at $W=1$~GeV and $W=1.76$~GeV, respectively.
The Dalitz-plot distributions largely depend on $\gamma$ both in shape
and magnitude.
Thus in these figures, we choose $\gamma$
where the Dalitz plot has the highest peak at the value of $W$.

In the right-hand sides of Figs.~\ref{fig:1000} and \ref{fig:1760}, the solid
curves (unitary model) are the distribution at a fixed
$m^2_{\pi^+\pi^-}$ of the
$x$-axis of the Dalitz plot on the left. 
The dotted curves (without $Z$) are obtained by turning off
the $Z$-diagram mechanism in our unitary calculations.
Clearly, the effect of the $Z$-diagram mechanism, which is the necessary
consequence of three-particle unitarity condition, is important, especially
in Fig.~\ref{fig:1000} where $a_1(1230)$ dominates.

For fitting the data with $\chi^2$-minimization, 
we need to assign an error to each data point.
We use the same error for all data points belonging to the same $W$.
For a given $W$, we use 5\% of the highest value in the Dalitz plot
distributions as the error.

We adjust the BW parameters $M_{\text{BW}}$, $\Gamma_{\text{BW}}$
and $\Lambda_{\text{BW}}$ in Eq.~(\ref{eq:BW}),
and all $M^*\to \pi R$ couplings constants $C_{(\pi R)_L,M^*}$ and 
cutoffs $\Lambda_{(\pi R)_L,M^*}$
to fit the Dalitz plots generated with our UM.
First we tried fitting with real $M^*\to \pi R$ couplings, but no
satisfactory fit is obtained. 
Thus we allow $M^*\to \pi R$ couplings to be complex, as usually done in the
IM analysis. 
To get high precision fits, we find that the IM needs to include
more  partial waves in the $M^*\rightarrow \pi R$ transition.
 This can be seen in Table~\ref{tab:pi2} in comparing the
parameters of the starting UM and IM.
The comparisons of the parameters for the other $M^*$ states are similar and therefore
need not to be given here.
(Note that the parameters of the UM are from selecting only few partial
waves which have information from the $^3P_0$ model.
 So the Table~\ref{tab:pi2} should not be used
in comparing the merits of each model. Namely,
if we start from the data generated from the IM with few partial waves, 
then the UM fits will also need
more partial waves. )

\begin{figure}[t]
\includegraphics[width=80mm]{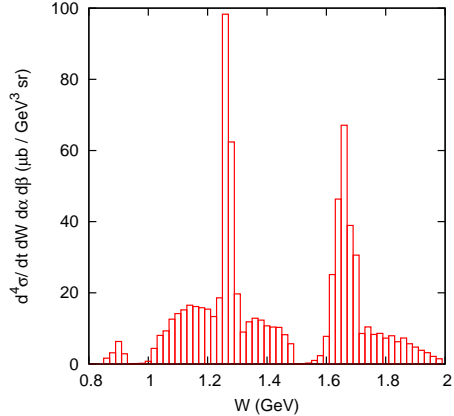}
\caption{\label{fig:bg} (color online)
The contribution from the constant background cross sections to the
integrated Dalitz plot obtained from Eq.~(\ref{eq:dalitz-unpol}) by
integrating over $m^2_{\pi^+\pi^-}$, $m^2_{\pi^+\pi^+}$ and $\gamma$.
}
\end{figure}

We find that even with much more partial waves 
in the fit, as seen in Table~\ref{tab:pi2},
the IM still cannot fit data well. The high precision fits
are obtained only when we 
add at each $W$ a flat and non-interfering background  to the 
Dalitz-plot distributions calculated from the IM.
The background contribution could in principle depend on kinematics and
interfere with the resonance contributions.
However, we follow the common practice of previous IM analyses, 
e.g., Refs.~\cite{e852,compass,clas}, to introduce the background contribution.
In Fig.~\ref{fig:bg}, 
we show the background contribution to the integrated Dalitz-plot distribution
[$m^2_{\pi^+\pi^-}$, $m^2_{\pi^+\pi^+}$ and $\gamma$ are integrated over
from Eq.~(\ref{eq:dalitz-unpol})]
which can be compared to Fig.~\ref{fig:w}.
The background contribution is highly $W$-dependent.
The largest contribution relative to the 
full Dalitz-plot distribution (the background plus the IM)
is at $W=1140$~MeV by 37\%.
Below the $a_2(1320)$ [$a_2(1700)$] peak, the contribution is 
22\% [29\%]. 
In some regions, on the other hand, 
the background contribution is almost zero as can be seen in the figure.

A good quality of fit has been obtained with the IM.
The IM gives Dalitz-plot distributions that are not 
distinguishable from the left-hand-side of Figs.~\ref{fig:1000} and \ref{fig:1760} of the UM.
The good fits can be seen from comparing
the solid (unitary model) and dashed (isobar fit) curves in the
right-hand-side of the same figures.

\subsection{$M^*$ Parameters}
\label{subsec:mstar}

\begin{table}[t]
\caption{\label{tab:pole} Pole positions $M^{\text{UM}}_R$ from the UM
 and $M^{\text{IM}}_R$ from the IM are compared. ${M^{\text{IM(BW)}}_R}$ are the
BW masses and widths $(M_{\text{BW}},-\Gamma_{\text{BW}}/2)$ from the IM. 
All are in unit of MeV.
}
\begin{ruledtabular}
\begin{tabular}[t]{ccccc}
$M^*$  & $M^0_{M^*}$& $M_R^{\text{UM}}$         & $M^{\text{IM}}_R$ & 
$M^{\text{IM(BW)}}_R$\\ \hline
$a_1$  & 1230	    & (913, $-$69)    & (940,$-$64)  &(1111,$-$600)       \\ 
       &  --	    & (1443, $-$342)  & (1201,$-$212)&(1391,$-$389)       \\ 
       & 1700	    & (1658,$-$53)    & (1672,$-$59) &(1676,$-$59)       \\ 
$a_2$  & 1320	    & (1263, $-$21)   & (1262,$-$22) &(1267,$-$24)       \\ 
       & 1700	    & (1652,$-$38)    & (1657,$-$48) &(1668,$-$52)       \\ 
$\pi_2$& 1670	    & (1786,$-$228)   & (1701,$-$220)&(1815,$-$283)       \\ 
       & 1800	    & (1722,$-$26)    & (1724,$-$34) &(1726,$-$34)       \\ 
$\pi_1$& 1600	    & (1599, $-$4)    &  (1599,$-$4)  &(1599,$-$4) 	        
 \end{tabular}
\end{ruledtabular}
\end{table}

We apply the analytic continuation method developed in Ref.~\cite{ssl-I}
 to extract the resonance
pole positions from searching for the solutions of
${\rm det}[G^{-1}_{M^*}(E)]=0$ of
Eq.~(\ref{eq:mstar-g1}) for our UM and
$[G^{BW}_{M^*}(E)]^{-1}=0$ of Eq.~(\ref{eq:BW}) for
the IM. The extracted poles $M_R$ for the UM and
$M^{IM}_R$ for IM are compared in Table~\ref{tab:pole}.
For the UM, we also list their bare masses $M^0_{M^*}$. 
For $a_1$, we see that two bare states
evolve into three resonance poles. The situation is similar to
$P_{11}$ nucleon resonances
 reported in Ref.~\cite{sjklms10}. 
In the last column, we also list
the BW positions $(M_{BW},-\Gamma_{BW}/2)$.

The results shown in Table~\ref{tab:pole} are 
similar to what we have observed in other analyses:
if the data are fitted equally well, the extracted resonance positions
are rather insensitive to the parametrization of the amplitudes as far as
the singularities of the scattering amplitudes like branch points
are far from the resonance pole.
Here we find a significant difference
between UM and IM in two poles associated with second $a_1$
(second row in Table~\ref{tab:pole}). 
The pole position of $a_1$ at $1443-i342$ MeV of the UM
is close to the branch points at
$1387-101 i$ MeV for the $\pi$-$f_2(1270)$ channel and $1487-167 i$ MeV
for the $\pi$-$f_0(1370)$ channel.
Thus the resonance shape of the amplitude at real $W$ are distorted by
those singularity at the complex energies.
On the other hand,
BW model has only a single complex energy
surface, and has no branch cut.
We have observed~\cite{sjklms10} the similar situation in the
study of Roper nucleon resonance $N^*(1440)$.

For $a_2$, we see from Fig.~\ref{fig:w} that two resonances for this partial
waves are well separated. Furthermore, $a_2(1320)$ is a very pronounced and
isolated resonance, like the $\Delta(1232)$ resonance. Thus
it is not surprising to see that the resonance positions for 
$a_2(1320)$ from two models
are almost identical and the BW value is also very close to the pole position.

For $\pi_2$, 
we assign the wide resonance as $\pi_2(1670)$ and the narrow one
as $\pi_2(1800)$ in Table~\ref{tab:pole}.
In UM, the mass of the wide resonance ($1786-228i$ MeV) is higher
than that of the narrow one ($1722-26i$ MeV), while
the order of two resonances is reversed in IM. 
Though IM reproduces well the data of Dalitz plot, the resonance positions for
overlapping resonances are sensitive to the reaction dynamics.

We next compare
the residues of the partial-wave amplitudes of $\gamma \pi \rightarrow R\pi$ transition
evaluated at the resonance  positions. 
To proceed, we first note that the total amplitude
of ${\gamma N\to \pi\pi\pi N^\prime}$ defined by 
Eqs.~(\ref{eq:gn-3pin})-(\ref{eq:decay-int}) can be cast into the following
form:
\begin{eqnarray}
T_{abc N^\prime,\gamma N} &=&
\sum^{\text{cyclic}}_{(a'b'c')}\sum_{R'}<a'b'c' N'|f_{a'b',R'}
\sum_{c''R''}G_{c'R',c''R''}T^{0}_{c''R''N', \gamma N}(W) \ ,
|\gamma N> \label{eq:tot-res}
\end{eqnarray}
with
\begin{eqnarray}
T^{0}_{c'R'N', \gamma N}(W) = 
\sum_\alpha\sum_{j_\alpha k_\alpha}\sum_{S^z_{M^*}}
\bar{\Gamma}_{c'R',M^*_{j_\alpha}} [G_{M^*}(W)]_{j_\alpha k_\alpha}
T_{M^*_{k_\alpha} N',\gamma N}(W) \ .
\label{eq:tres}
\end{eqnarray}
By using Eq.~(\ref{eq:opep}) for $T_{M^*N',\gamma N}(W)$ and performing the
partial-wave expansion, 
the following partial wave amplitude 
$T^{\alpha}_{(\gamma\pi)_{L_i}\to (R^{LI}_i\pi)_{L_f}} (W)$
 can be separated from 
$T^{0}_{c'R'N', \gamma N}(W)$:
\begin{eqnarray}
\label{eq:res-amp}
T^{\alpha}_{(\gamma\pi)_{L_i}\to (R^{LI}_i\pi)_{L_f}} (W)&=&
\sum_{j_\alpha k_\alpha}
 \bar\Gamma_{(R^{LI}_i\pi)_{L_f},M_{j_\alpha}^*} (W,p_{R^{LI}_i})
\nonumber\\
&\times&
\left[G_{M^*} (W) \right]_{j_\alpha k_\alpha}
\left({e\over g_\rho} \sum_{\rho'} \sqrt{E_{\rho'}(p_{\rho'})\over
 E_\gamma(p_{\rho'})}  {\Gamma}_{M_{k_\alpha}^*, (\rho' \pi)_{L_i}} (p_{\rho'})\right) \ ,
\label{eq:res-res}
\end{eqnarray}
where $L_i$ ($L_f$) is the orbital angular momentum of $\gamma\pi$
($R^{LI}_i\pi$).
The symbol $p_{R^{LI}_i}$
($p_{\rho'}$) denotes the on-shell momentum of ${R^{LI}_i}$ ($\rho'$)
that satisfies
$W=M_R=E_{R^{LI}_i}(p_{R^{LI}_i})+E_\pi(p_{R^{LI}_i})$ where
the bare mass of ${R^{LI}_i}$ in our model is used to calculate the
energy $E_{R^{LI}_i}(p_{R^{LI}_i})$.

The residue at each pole position $M_R$ is then defined by a contour integration along a 
closed path ($C_{M_R}$) around the pole position $M_R$. We further
multiply it by phase-space factors to define
\begin{eqnarray}
\label{eq:residue}
B^{\alpha, M_R}\left((\gamma\pi)_{L_i}\to (R^{LI}_i\pi)_{L_f}\right) &=& -\sqrt{\rho_{\gamma\pi}(M_R)\
 \rho_{R^{LI}_i\pi}(M_R)} 
\oint_{C_{M_R}} \!\!\! d\bar W\ 
T^{\alpha}_{(\gamma\pi)_{L_i}\to (R^{LI}_i\pi)_{L_f}} (\bar W)
\ ,
\end{eqnarray}
where the phase space factors are 
\begin{eqnarray}
\rho_{R\pi}(W) &=& \pi{p_R E_R(p_R) E_\pi(p_R)\over W} \ , \nonumber \\
\rho_{\gamma\pi}(W) &=& \pi \frac{p^2_\gamma E_\pi(p_\gamma)}{W}.
\end{eqnarray}

For the IM, we can cast Eq.~(\ref{eq:gn-3pin-im}) into
the same form of Eqs.~(\ref{eq:tot-res})-(\ref{eq:res-res}). It is
straightforward to see that the resulting
$\gamma \pi \rightarrow \pi R$ amplitude is 
\begin{eqnarray}
T^{\alpha(\text{IM})}_{(\gamma\pi)_{L_i}\to (R^{LI}_i\pi)_{L_f}} (W)&=&
\sum_{j_\alpha}
 \Gamma_{(R^{LI}_i\pi)_{L_f},M^*_{j_\alpha}} (p_{R^{LI}_i})
\nonumber\\
&\times&
[G^{BW}_{M^*} (W)]_{j_\alpha j_\alpha} 
\left({e\over g_\rho} \sum_{\rho'} \sqrt{E_{\rho'}(p_{\rho'})\over
 E_\gamma(p_{\rho'})}  \Gamma_{M^*_{j_\alpha}, (\rho' \pi)_{L_i}} (p_{\rho'})\right) \ ,
\label{eq:res-res-im}
\end{eqnarray}
where $G^{BW}_{M^*} (W)$ is given in Eq.~(\ref{eq:BW}). 
Its residue can be measured by using  Eq.~(\ref{eq:residue})
with the replacement 
$T^{\alpha}_{(\gamma\pi)_{L_i}\to (R^{LI}_i\pi)_{L_f}} (W)\rightarrow T^{\alpha(\text{IM})}_{(\gamma\pi)_{L_i}\to (R^{LI}_i\pi)_{L_f}} (W)$.

Following the usual procedure, the residue of
the BW parametrization at $W=M_{\text{BW}}$ is then
obtained from Eq.~(\ref{eq:res-res-im}) by replacing 
$G^{\text{BW}}_{M^*}(W)$ with $1/2$.
Multiplying the same phase factors, we then define
for the BW: 
\begin{eqnarray}
\label{eq:im_bw}
B^{\alpha, M^{j_\alpha}_{\text{BW}}}_{\rm IM(BW)}\left((\gamma\pi)_{L_i}\to (R^{LI}_i\pi)_{L_f}\right) 
&=& -{1\over 2}\sqrt{\rho_{\gamma\pi}(M^{j_\alpha}_{\text{BW}})\
 \rho_{R^{LI}_i\pi}(M^{j_\alpha}_{\text{BW}})}\ 
 \Gamma_{(R^{LI}_i\pi)_{L_f},M^*_{j_\alpha}} (p_{R^{LI}_i})
\nonumber\\
&\times& 
\left({e\over g_\rho} \sum_{\rho'} \sqrt{E_{\rho'}(p_{\rho'})\over
 E_\gamma(p_{\rho'})}  \Gamma_{M^*_{j_\alpha}, (\rho' \pi)_{L_i}} (p_{\rho'})\right) \ ,
\end{eqnarray}
where the on-shell momenta are for the BW mass ($M_{\text{BW}}$)
taken as the total energy.

We have determined the residues for $M^*$s of UM and IM using
Eq.~(\ref{eq:residue}), 
and also the conventional BW residues for IM using Eq.~(\ref{eq:im_bw}).
In Table~\ref{tab:coup} we present only results which are useful in 
revealing the
essential differences between these three residues.
Because the overall phase is arbitrary, we choose the overall phase for IM such
that the phases of $B(\gamma\pi\to R^{11}_1\pi)$
for the most prominent $a_2(1320)$ are the same for all three cases in this table.
Guided by the results of the $W$-dependence of cross sections
shown in Fig.~\ref{fig:w} and the pole positions shown in
Table~\ref{tab:pole}, we focus on three different situations:
(a) the resonances are broad such as the second $a_1(1230)$;
(b) resonances in the same partial wave are narrow and isolated  such as 
$a_2(1320)$ and $a_2(1700)$;
(c) resonances in the same partial wave overlap such as  $\pi_2(1670)$ and
$\pi_2(1800)$.

\begin{figure}
\includegraphics[width=80mm]{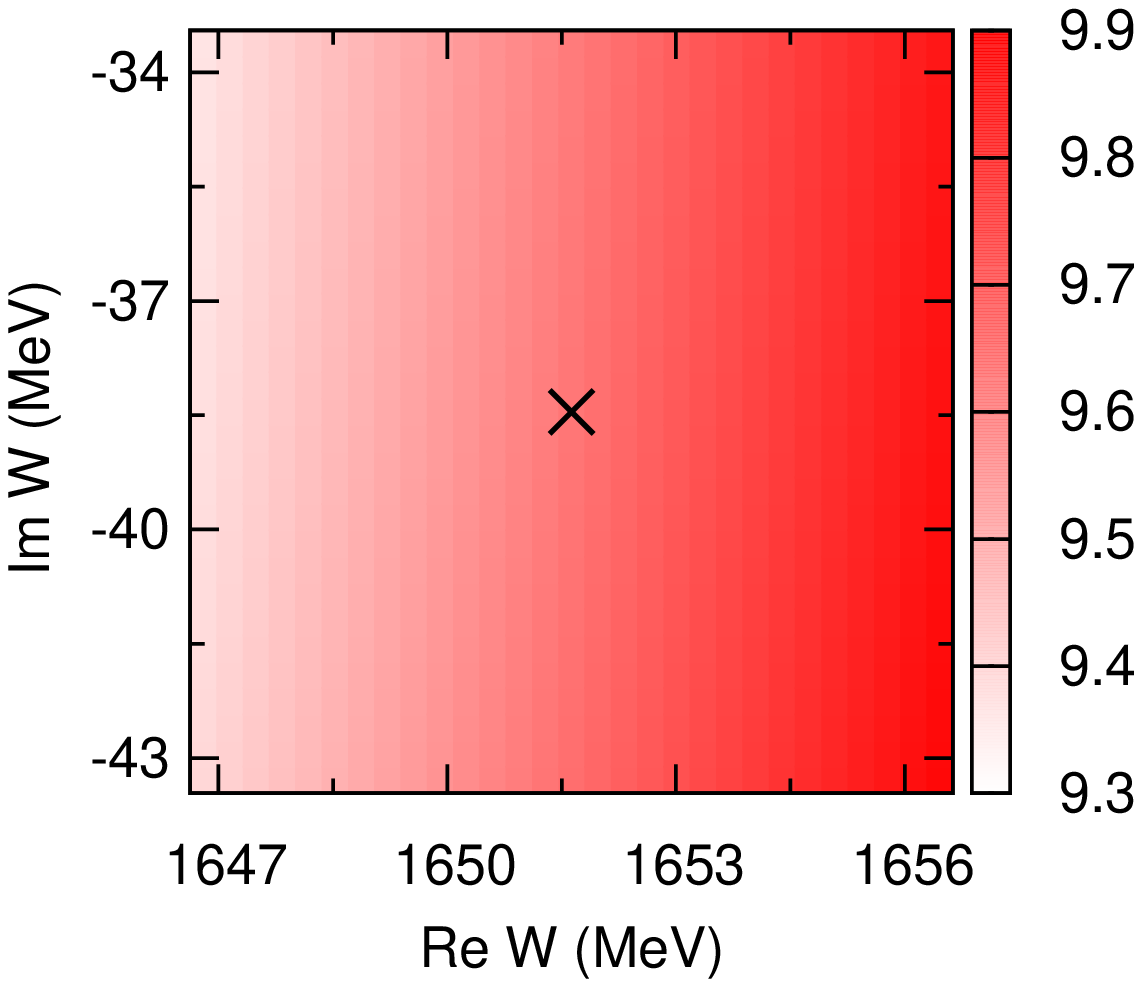}
\includegraphics[width=80mm]{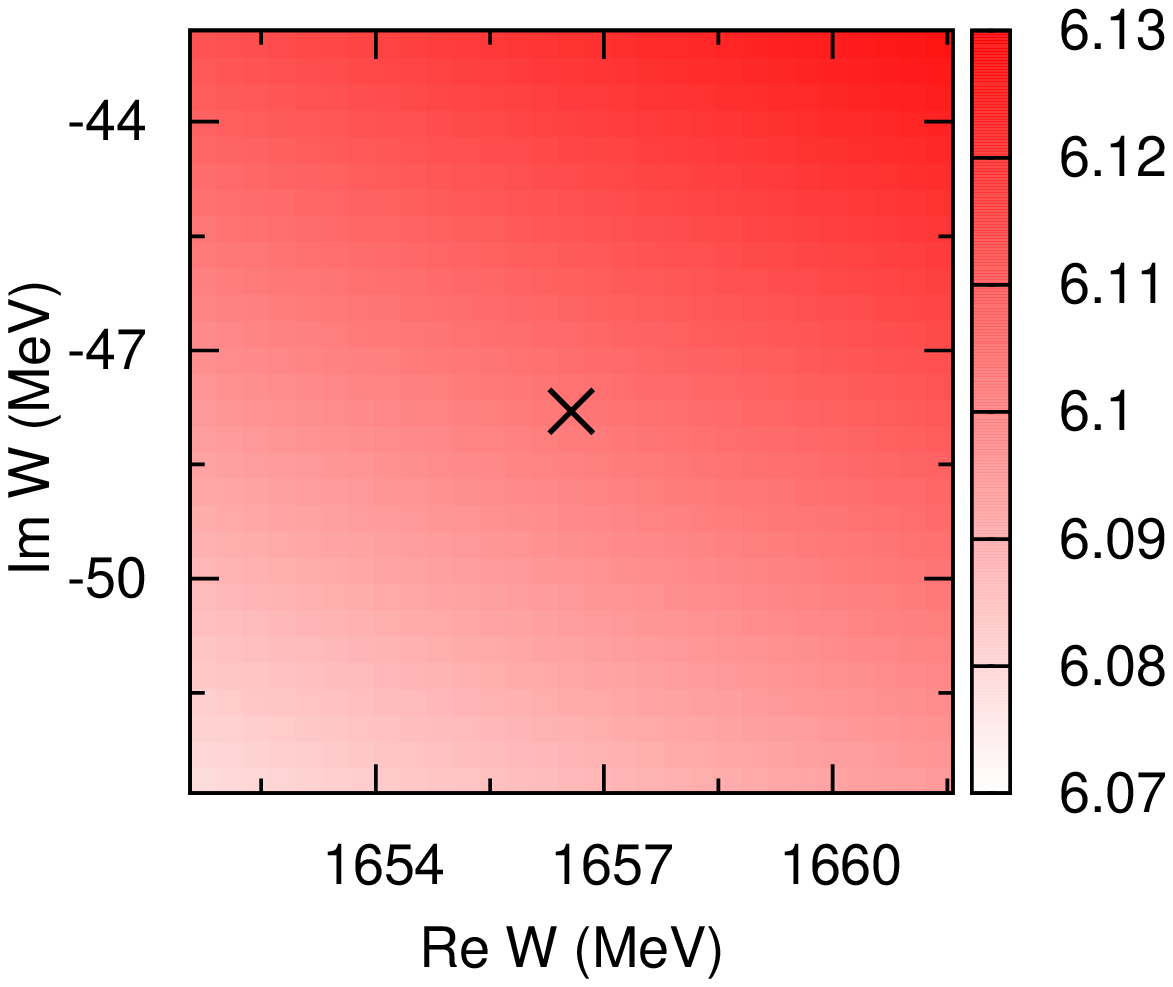}
\caption{\label{fig:contour}
(color online)
The real part of
 $(W-M_R) T^{\alpha}_{(\gamma\pi)_2\to (R^{11}_1\pi)_2}(W)$ 
[10$^{-6}$ MeV$^{-1/2}$]
for $a_2(1700)$  from the UM
 (left), and from the isobar plus BW model (right).
The pole position for $a_2(1700)$ is indicated by the cross.
The resonant amplitude, $T^{\alpha}_{(\gamma\pi)_2\to (R^{11}_1\pi)_2}(W)$, is 
 defined in Eq.~(\ref{eq:res-amp}).
}
\end{figure}

We see that
for the broad resonance 2nd-$a_1(1230)$, the extracted residues are 
drastically different between UM and IM. 
This is of course partly due to the difficulties
in getting the same pole position 
for the reason discussed in this subsection.
For the $a_2(1320)$ that gives the very pronounced peak in Fig.~\ref{fig:w},
both the resonance positions and residues agree almost perfectly between
UM and IM.
This is not surprising since it is similar to the situation of the well
known $\Delta(1232)$
resonance in the $\pi N$ scattering. For the $a_2(1700)$, there is some
difference in resonance positions and residues. Since the residues
from UM and IM are from loop integration around the pole position as 
given in Eq.~(\ref{eq:residue}), their differences originate from the
differences in their amplitudes near the pole position. 
This is illustrated in Fig.~\ref{fig:contour} for this resonance. 
We see that the amplitudes from UM and IM are rather different near their pole positions and
hence lead to the differences in residues which are calculated from loop integration
of the amplitude around the pole position.
This indicates that the resonance properties extracted from data are 
not independent
of analysis method, and thus it is essential to have a parametrization
of the amplitude with theoretical constraint.
Finally, for $\pi_2(1670)$ and $\pi_2(1800)$, it is not surprising to see
that the phases of their residues from UM and IM are very different 
since two resonances are overlapping.

\begin{table}[t]
\caption{\label{tab:coup} 
The pole positions $M_R$ and the weighted-residues $B(\gamma \pi \rightarrow \pi R^{LI}_i)$ 
defined in Eq.~(\ref{eq:residue}) from UM and IM fit are
compared.
$R^{LI}_i$ denotes the $i$-th bare $R$ state
with the spin $L$ and the isospin $I$.
For $M_R$, they are listed as $({\rm Re}(M_R), {\rm Im}(M_R))$ [MeV].
The column under IM(BW) are from IM but using the usual BW procedure
with their masses listed as $M_R =(M_{BW}, -\Gamma_{BW}/2$).
For the weighted-residues, we list as ($|B|$ [MeV],  $\phi$ [deg]) of the expression
 $B(\gamma \pi \rightarrow \pi R)=|B|e^{i\phi}$.
The overall phase for IM is chosen such that the phases of $B(\gamma\pi\to R^{11}_1\pi)$
for the most prominent $a_2(1320)$ are the same as UM.
}
\begin{ruledtabular}
\begin{tabular}[t]{ccccccc}
$M^*$ & &                         &$(L_i,L_f)$&       UM & IM (pole) & IM (BW) \\\hline
2nd-$a_1$(1230)      & $M_R$&                    &     & $(1443 , -342)$ & $(1201 , -212)$ & $(1391,-389)$\\
      & &$B(\gamma\pi\to\pi R^{11}_1)$		 &(0,0)&$ 64.1,\ -67.$&$ 28.4,\ -171.$&$ 33.6,\ -150.$\\
      & &$B(\gamma\pi\to\pi R^{11}_1)$		 &(0,2)&$  8.1,\  34.$&$  0.0,\  180.$&$  0.0,\  180.$\\
\hline
$a_2(1320)$ & $M_R$&                    &     & $(1263 ,  -21)$ &$(1262 ,  -22)$ & $(1267,-24) $\\
      & &$B(\gamma\pi\to\pi R^{11}_1)$		 &(2,2)&$  6.2,\ 171.$&$  6.2,\ 171.$&$  6.7,\ 171.$\\
$a_2(1700)$      & $M_R$&                    &     & $(1652 ,  -38)$ & $(1657 ,  -48)$ & $(1668,-52) $\\
      & &$B(\gamma\pi\to\pi R^{11}_1)$		 &(2,2)&$  9.4,\ 147.$&$  6.5,\ 139.$&$  7.1,\ 140.$\\
\hline
$\pi_2$(1670) & $M_R$&                    &     & $(1785 , -229)$ & $(1701 , -220)$ & $(1815,-283)$\\
      & &$B(\gamma\pi\to\pi R^{00}_1)$		 &(1,2)&$  2.0,\ -123.$&$  4.5,\  129.$&$  5.4,\  176.$\\
      & &$B(\gamma\pi\to\pi R^{11}_1)$		 &(1,1)&$ 31.8,\ -134.$&$ 37.6,\  167.$&$ 38.6,\  168.$\\
      & &$B(\gamma\pi\to\pi R^{11}_1)$		 &(1,3)&$  1.9,\ -101.$&$  0.6,\  153.$&$  0.7,\  176.$\\
$\pi_2$(1800)        & $M_R$&                    &     & $(1722 ,  -26)$ & $(1724 ,  -34)$ & $(1726,-34) $\\
      & &$B(\gamma\pi\to\pi R^{00}_1)$		 &(1,2)&$  0.6,\   73.$&$  0.6,\  -26.$&$  0.6,\  -35.$\\
      & &$B(\gamma\pi\to\pi R^{11}_1)$		 &(1,1)&$  4.5,\    4.$&$  4.3,\   -7.$&$  4.3,\  -21.$\\
      & &$B(\gamma\pi\to\pi R^{11}_1)$		 &(1,3)&$  0.2,\   14.$&$  0.0,\  180.$&$  0.0,\  180.$\\
 \end{tabular}
\end{ruledtabular}
\end{table}

\section{Summary}
\label{sec:summary}

We applied the unitary coupled-channels model developed in
Ref.~\cite{3pi} to investigate the issues concerning the extraction of
meson resonances from the three-pions photoproduction reaction on the nucleon.
Our aim here is to examine the importance of the three-body unitarity, which is not
accounted for rigorously
in the commonly used IM analyses for the resonance extraction.
This has been done by comparing the resonance parameters extracted with
our UM and  an IM both of which
reproduce the same Dalitz plots over relevant kinematical region. 
We also compare the resonance parameters with the usual BW parameters of
the same IM.

We found that the good IM fits to the Dalitz plots generated from 
the UM can be achieved only when the $M^*\to \pi R$ coupling are allowed to
become complex and a flat background is added at each $W$.
The resonance positions from the two models agree well, except for the resonance 
poles whose tails are partly blocked by branch cuts on the Riemann surface 
from reaching at the physical real $W$ axis
and for the overlapping resonances.
The residues of the resonant amplitudes
extracted from the two models and those from the usual BW procedure  
agree well only for the isolated resonances with narrow
widths. Most of the extracted residues
for overlapping resonances could be drastically different.
Our results suggest that even with high precision data, the resonance extraction
should be based on models within which the amplitude parametrization is
constrained by three-particle unitarity condition.

\acknowledgments
SXN is the Yukawa Fellow and his work is supported in part by Yukawa Memorial Foundation, the
Yukawa International Program for Quark-hadron Sciences (YIPQS), and by Grants-in-Aid
for the global COE program ``The Next Generation of Physics, Spun from Universality and
Emergence'' from MEXT. 
SXN was also supported
by the U.S. Department of Energy, Office of Nuclear Physics Division, under Contract No.
DE-AC05-06OR23177 under which Jefferson Science Associates operates
Jefferson Lab.
HK acknowledges the support by the HPCI Strategic Program (Field 5 ``The
Origin of Matter and the Universe'') of Ministry of Education, Culture,
Sports, Science and Technology (MEXT) of Japan. 
TS is supported by JSPS KAKENHI (Grant Number 24540273).
This work is also supported by the U.S. Department of Energy, Office of Nuclear Physics
Division, under Contract No. DE-AC02-06CH11357. This work used resources of the
National Energy Research Scientific Computing Center, which is supported by the Office of
Science of the U.S. Department of Energy under Contract No. DE-AC02-05CH11231, and
resources provided on ``Fusion'', a 320-node computing cluster operated by the Laboratory
Computing Resource Center at Argonne National Laboratory.


\begin{thebibliography}{99}

\bibitem{klempt}
E. Klempt and A. Zaitsev, Phys. Rep. {\bf 454}, 1 (2007).

\bibitem{e852}
G. S. Adams et al. (E852 Collaboration), Phys. Rev. Lett. {\bf 81}, 5760 (1998);\\
S. U. Chung et al. (E852 Collaboration), Phys. Rev. D {\bf 65}, 072001 (2002);\\
A. R. Dzierba et al., Phys. Rev. D {\bf 73}, 072001 (2006).

\bibitem{compass}
M. Alekseev et al. (COMPASS Collaboration), Phys. Rev.
Lett. {\bf 104}, 241803 (2010).

\bibitem{clas} M. Nozar et al. (CLAS Collaboration), Phys. Rev. Lett.
{\bf 102}, 102002 (2009).

\bibitem{gluex}
D. S. Carman (The GlueX Collaboration), in Hadron
Spectroscopy: Eleventh International Conference on
Hadron Spectroscopy, edited by A. Reis, C. G\"obel, J. de
S\'a Borges, and J. Magnin, AIP Conf. Proc. No. 814 (AIP,
New York, 2006).

\bibitem{3p0}
T. Barnes, F. E. Close, P. R. Page, and E. S. Swanson, 
Phys. Rev. D {\bf 55}, 4157 (1997).

\bibitem{3pi}
H. Kamano, S. X. Nakamura, T.-S. H. Lee, and T. Sato, 
Phys. Rev. D {\bf 84}, 114019 (2011).

\bibitem{kuhn}
J. H. K\"uhn and E. Mirkes, Z. Phys. C {\bf 56}, 661 (1992).

\bibitem{chung}
S. U. Chung, BNL PREPRINT 76975-2006-IR.

\bibitem{sl96}
T.~Sato and T.-S.~H. Lee, Phys. Rev. C {\bf 54}, 2660 (1996).

\bibitem{msl}
A. Matsuyama, T. Sato, and T.-S. H. Lee, Phys. Rep. {\bf 439}, 193 (2007).

\bibitem{IKP}
N. Isgur, R. Kokoski, and J. Paton, Phys. Rev. Lett {\bf 54}, 869 (1985).

\bibitem{ssl-I}
N. Suzuki, T. Sato, and T. S. H. Lee, Phys. Rev. C {\bf 79}, 025205 (2009).

\bibitem{sjklms10}
N. Suzuki, B. Juli\'a-D\'{\i}az, H. Kamano, T.-S. H. Lee, A. Matsuyama, and T. Sato,
Phys. Rev. Lett. {\bf 104}, 042302 (2010). 

\end{thebibliography}
\end{document}